\documentclass[11pt,twoside]{article}
\usepackage[pdftex]{graphicx}
\usepackage{amsmath}
\usepackage{amssymb}
\usepackage{cite}
\usepackage{dsfont}

\begin{document}
\newcommand{\pst}{\hspace*{1.5em}}

\newcommand{\rigmark}{\em Journal of Russian Laser Research}
\newcommand{\lemark}{\em Volume 30, Number 5, 2009}

%\lhead[\fancyplain{\rigmark, {\em \lemark}}{\rigmark}]{\fancyplain{\rigmark, {\em \lemark}}{\lemark}}
%\chead{}\rhead[\fancyplain{}{\lemark}]{\fancyplain{}{\rigmark}}
%\plainfootrulewidth 0.4pt
\newcommand{\be}{\begin{equation}}
\newcommand{\ee}{\end{equation}}
\newcommand{\bm}{\boldmath}
\newcommand{\ds}{\displaystyle}
\newcommand{\bea}{\begin{eqnarray}}
\newcommand{\eea}{\end{eqnarray}}
\newcommand{\ba}{\begin{array}}
\newcommand{\ea}{\end{array}}
\newcommand{\arcsinh}{\mathop{\rm arcsinh}\nolimits}
\newcommand{\arctanh}{\mathop{\rm arctanh}\nolimits}
\newcommand{\bc}{\begin{center}}
\newcommand{\ec}{\end{center}}

\newcommand{\braket}[2]{\langle{#1}|{#2}\rangle}
\newcommand{\ketbra}[2]{|{#1}\rangle\langle{#2}|}
\newcommand* {\bra}[1]{\ensuremath{\langle {#1} |}}
\newcommand* {\ket}[1]{\ensuremath{| {#1} \rangle}}

\thispagestyle{plain}

\label{sh}

%\lfoot[\fancyplain{\ \\[1mm] \thepage}{\ \\[1mm]\thepage}]{\fancyplain{}{}}

\begin{center} {\Large \bf
\begin{tabular}{c}
SENSITIVITY TO INITIAL NOISE IN
\\[-1mm]
MEASUREMENT-INDUCED NONLINEAR QUANTUM DYNAMICS
\end{tabular}
 } \end{center}

\bigskip

\bigskip

\begin{center} {\bf
Orsolya K\'{a}lm\'{a}n$^{1*}$, Tam\'{a}s Kiss$^1$ and Igor Jex$^2$
}\end{center}

\medskip

\begin{center}
{\it
$^1$Institute for Solid State Physics and Optics, Wigner Research Centre, Hungarian Academy of Sciences, 
P.O. Box 49, H-1525
Budapest, Hungary}

\smallskip
{\it
$^2$Faculty of Nuclear Sciences and Physical Engineering, Czech Technical University in Prague \\
B\v rehov\'a 7, 115 19 Praha 1 - Star\'e M\v esto, Czech Republic} \smallskip

$^*$Corresponding author e-mail:~~~kalman.orsolya~@~wigner.mta.hu\\
\end{center}

\begin{abstract}\noindent
We consider a special iterated quantum protocol with measurement-induced nonlinearity for qubits, where all pure initial states on the Bloch sphere can be considered chaotic. The dynamics is ergodic with no attractive fixed cycles. We show that initial noise radically changes this behavior. The completely mixed state is an attractive fixed point of the dynamics induced by the protocol. Our numerical simulations strongly indicate that initially mixed states all converge to the completely mixed state. The presented protocol is an example, where gaining information from measurements and employing it to control an ensemble of quantum systems enables us to create ergodicity, which in turn is destroyed by any initial noise.
\end{abstract}

\medskip


\section{Introduction}
\pst

J\'{o}zsef Janszky's interest was attracted towards quantum optics by studying the interplay of nonlinear dynamics and noise in crystals. We share his enthusiasm for nonlinearities as they can be the source of many interesting effects. We consider here the interplay of nonlinear dynamics and noise within a scheme motivated by iterated quantum information protocols. 

Ergodicity in quantum physics usually refers to systems featuring quantum chaos, where the corresponding classical system is chaotic and the energy eigenstates spread all over the phase space. In such closed systems, however, two initially close pure quantum states will remain close throughout the dynamics\cite{chaosbook}. Only in open quantum systems it is possible that initial states evolve with decreasing overlap.

In quantum information theory, the combination of unitary evolution, postselection based on measurement results, and subsequent further manipulation provides a useful tool, e.g. for entanglement distillation. An entangling gate and a subsequent measurement applied on a pair of qubits leads to a nonlinear transformation as originally suggested by Bechmann-Pasquinucci, Huttner and Gisin\cite{Gisin}. Repeating the above procedure can lead to entanglement distillation \cite{Bennett1996,Bennett1996b,Alber2001}.  The same scheme acting pairwise on an ensemble of identically prepared systems in a pure quantum state may result in a postselected ensemble in a nonlinearly transformed pure state, providing a possible physical realization of nonlinear quantum channels \cite{Manko2014,Manko2015}. Iterating such nonlinear quantum state transformations may result in strong dependence on the initial conditions and in complex chaos \cite{Kiss2006}. Chaotic evolution of pure quantum states has consequences for the time-dependence of entanglement as well, in the case of applying the iterative dynamics on systems consisting of qubit pairs. Entanglement itself can evolve chaotically \cite{Kiss2011}.

The above described iterative dynamics of qubits is effectively described by an iterated quadratic rational map on complex numbers. In turn, given any quadratic (or higher order) rational map on complex numbers, one can construct a quantum gate and a corresponding protocol, which realizes the map with qubits \cite{Gilyen2016}. Representing pure initial states on the Bloch sphere, it can be divided into two complementary sets: initial states from the Fatou set converge to an attractive fixed cycle of the dynamics, while the remaining states, forming the Julia set, are considered chaotic in the mathematical literature \cite{Milnorbook}. The Julia set is a fractal in most cases, with measure zero on the surface of the Bloch sphere. In an exceptionally interesting case, the whole Bloch sphere forms the Julia set and the Fatou set is empty. For such cases the corresponding so-called Latt\`{e}s-type of map has no attractive fixed cycle and all initial states are chaotic. Such a map can be ergodic int the sense that it is exponentially mixing for pure initial states, which means tiny uncertainty about the initial state evolves exponentially fast to a complete uncertainty \cite{Gilyen2016}.

In this paper, we study how initially present noise affects the dynamics. We show that in this case there is no purification, all noisy initial states will eventually converge to the completely mixed state. This implies that arbitrarily small initial noise will destroy ergodicity.

The paper is organized as follows. In Sec.~2 we introduce a Latt\`{e}s-type of nonlinear quantum protocol. We give its generalization for noisy (i.e., mixed) inputs in Sec.~3. In Sec.~4 we analyze the dynamics for initially noisy inputs and show that any infinitesimally small initial noise in the input state eventually destroys the chaotic behaviour as the output will converge to the maximally mixed state. 
We conclude in Sec.~5.

\section{Ergodic nonlinear map for pure initial states}\label{nonlinmap}
\pst
We consider the following quantum protocol. As inputs of a CNOT quantum gate \cite{NCH} let us have two independent identical copies of the pure qubit state 
\be
\ket{\psi_{0}}=\frac{\ket{0}+z\ket{1}}{\sqrt{1+\left|z\right|^{2}}}, \qquad z\in \hat{\mathbb{C}}=\mathbb{C}\cup \infty.
\ee 
It can be easily seen that after the CNOT operation, the resulting two-qubit state is given by
\be
U_{\mathrm{CNOT}}\left(\ket{\psi_{0}}\otimes\ket{\psi_{0}}\right)=\frac{1}{1+\left|z\right|^{2}}\left(\ket{00}+z\ket{01}+z\ket{10}+z^{2}\ket{11}\right)
\ee
If we make a projective measurement on the target qubit and the result is $0$, then the state of the other qubit is given by
\be
\ket{\psi_{1}}=\left(\mathds{1} \otimes \ketbra{0}{0}\right)U_{\mathrm{CNOT}}\left(\ket{\psi_{0}}\otimes\ket{\psi_{0}}\right)=\frac{\ket{0}+z^{2}\ket{1}}{\sqrt{1+\left|z\right|^{4}}}.
\ee 
(If the measurement result is $1$ we discard the control qubit). Thus, in possession of the measurement result being $0$, we can know that the state of the remaining qubit is transformed in a nonlinear way by the map $f_{0}(z)=z^{2}$. 

If we have more copies of the same initial state $\ket{\psi_{0}}$, i.e., an ensemble, then this protocol may be iterated by forming new pairs of the remaining qubits and repeating the protocol on them. The above quantum protocol is the simplest one which results in the nonlinear transformation of the initial qubit state. However, one can think of augmenting this protocol by a single-qubit unitary transformation $U$ in each step, i.e., before the next application of the CNOT gate and the post-selection, we can transform $\ket{\psi_{n}}$ into $U\ket{\psi_{n}}$. It turns out that this leads to a large variety of nonlinear maps (more precisely complex quadratic rational maps) which, when iterated, leads to a very rich variety of dynamics \cite{Gilyen2016}, where sets of states which evolve chaotically may appear. 

In the special case when the single-qubit unitary is 
\be
U_{\mathrm{L}}=\frac{1}{\sqrt{2}}\left(\begin{array}{cc} 1 & i \\ i & 1 \end{array}\right),
\ee
then the corresponding nonlinear map is given by 
\be
f_{\mathrm{L}}(z)=\frac{z^{2}+i}{iz^{2}+1}.
\label{f_L}
\ee 
The fixed cycles of $f_{\mathrm{L}}$ can be determined from the condition $f_{\mathrm{L}}^{\circ n}(z)=z$. In Table~\ref{table} the analytically determined fixed cycles are shown up to length two. It can be shown that all of these cycles are repelling.
\smallskip

%%%%%%%%%%%%%%%%%%%%%%%%%%%%%%%%%%%%%%%%%%%%%%%%%%%%%%%%%%%%%%%%%%%%%%%%%%%%%%%%
\begin{table}[h!]
\centering
\begin{tabular}{|c|c|c|}%{ |m{0.5cm}|p{5.0cm}|m{2.2cm}| }
 \hline
 \multicolumn{2}{|c|}{Fixed cycles up to length $2$ of $f_{\mathrm{L}}$} \\
 \hline
 $c_{1}$ & $1$ \\
 \hline
 \rule{0pt}{4ex}
 $c_{2}$ & $\sqrt{\dfrac{i-2}{2}}-\dfrac{i+1}{2}$
 %\vspace{0.2ex} 
 \\
 \hline
 \rule{0pt}{4ex}
 $c_{3}$ & $-\sqrt{\dfrac{i-2}{2}}-\dfrac{i+1}{2}$ 
%\vspace{0.2ex}
\\
\hline
 \rule{0pt}{4ex}
 $c_{4}$ & $\sqrt{\dfrac{-i-2}{2}}+\dfrac{i-1}{2} \leftrightarrow -\sqrt{\dfrac{-i-2}{2}}+\dfrac{i-1}{2}$ 
%\vspace{0.2ex}
\\
\hline
\end{tabular}
\caption{Fixed cycles up to length two of the map $f_{\mathrm{L}}$.}
\label{table}
\end{table}
%%%%%%%%%%%%%%%%%%%%%%%%%%%%%%%%%%%%%%%%%%%%%%%%%%%%%%%%%%%%%%%%%%%%%%%%%%%%%%%%

$f_{\mathrm{L}}$ is one of a few special so-called Latt\`{e}s maps \cite{Milnor2006} and as such has gained a
lot of attention in the theory of complex dynamical systems \cite{Milnorbook}. Its peculiarity lies in the fact that it does not have any attractive fixed cycles, therefore, when iterated, the resulting values do not converge for any initial $z$, or with other words, its so-called Fatou set is empty \cite{Milnorbook}. On the other hand, its so-called Julia set \cite{Milnorbook} (which is the closure of its repelling fixed cycles) is identical to $\mathbb C$. Since the points of the Julia set are known to exhibit chaotic dynamics, this means that all initial (pure) qubit states will evolve chaotically. Moreover, it has been shown that such a nonlinear quantum protocol has exponential sensitivity to the initial conditions and can act as the so-called ''Schr\"{o}dinger's microscope'' \cite{Lloyd2000}. 

\section{Generalization of the nonlinear map for noisy initial states}\label{gen}
\pst

In order to investigate the effect of initial noise on the dynamics, we need to generalize the transformation $f_{\mathrm{L}}$ for mixed-state inputs. 

Let us consider two independent, identical qubits each in the mixed initial state 
\be
\rho_{0}=\frac{1}{\rho_{11}+\rho_{22}}\left(\begin{array}{cc} \rho_{11} & \rho_{12} \\ \rho_{12}^{\ast} & \rho_{22} \end{array} \right).
\ee
Then, it can be easily shown that after the application of the CNOT gate and the projection of the target qubit to $\ket{0}$, the control qubit will be left in the state
\be
\rho_{1}=\frac{1}{\rho_{11}^{2}+\rho_{22}^{2}}\left(\begin{array}{cc} \rho_{11}^{2} & \rho_{12}^{2} \\ \left(\rho_{12}^{\ast}\right)^{2} & \rho_{22}^{2} \end{array} \right).
\ee
It can be seen that the elements of $\rho_{0}$ are squared (apart from normalization) due to this elementary nonlinear protocol. Let us denote the squaring operation by $S$ (i.e., $\rho_{1}=S\left(\rho_{0}\right)$). If, similarly to the pure-state case, we augment the protocol with the single-qubit unitary $U_{\mathrm{L}}$ after $S$ in each step, then the resulting density matrix after the $n$th step will be
\be
\rho_{n}=\left(U_{\mathrm{L}}S\left(\rho_{0}\right)U_{\mathrm{L}}^{\dagger}\right)^{\circ n}=\left({\cal M}_{\mathrm{L}}\!\left(\rho_{0}\right)\right)^{\circ n},
\ee
where ${\circ n}$ means that the operation is executed $n$ times, and ${\cal M}_{\mathrm{L}}$ denotes the overall transformation of one step of the protocol. 

Let us parametrize $\rho_{0}$ with its Bloch-sphere coordinates $u,v,w \in \mathbb{R}$ as
\be
\rho_{0}=\frac{1}{2}\left(\mathds{1}+u\sigma_{x}+v\sigma_{y}+w\sigma_{z}\right)=\frac{1}{2}\left(\begin{array}{cc} 1+w & u-iv \\ u+iv & 1-w \end{array} \right),
\ee 
where $\sigma_{i}$ ($i=x,y,z$) are the Pauli matrices, and $u^{2}+v^{2}+w^{2}\leq 1$, and the purity of $\rho_{0}$ is given by $P=\left(u^{2}+v^{2}+w^{2}+1\right)/2$. Then, after one step of the protocol, we find that the map ${\cal M}_{\mathrm{L}}$ transforms $(u,v,w)$ into
\be
\mathcal{U}=\frac{u^{2}-v^{2}}{1+w^{2}}, \quad \mathcal{V}=\frac{2w}{1+w^{2}}, \quad \mathcal{W}=-\frac{2uv}{1+w^{2}}.
\label{uvw_transf}
\ee  

%%%%%%%%%%%%%%%%%%%%%%%%%%%%%%%%%%%%%%%%%%%%%%%%%%%%%%%%%%%%%%%%%%%%%%%%%%%%%%%%
\begin{table}[h!]
\centering
\begin{tabular}{|c|c|c|}%{ |m{0.5cm}|p{5.0cm}|m{2.2cm}| }
 \hline
 \multicolumn{2}{|c|}{Bloch-sphere coordinates of fixed cycles up to length $2$ of $\mathcal{M}_{\mathrm{L}}$} \\
 \hline
 $C_{0}$ & $\left(0,0,0\right)$ \\
 \hline
 $C_{1}$ & $\left(1, 0, 0\right)$ \\
 \hline
 $C_{2}$ & $\left(-0.382,0.786,0.486\right)$ \\
 \hline
 $C_{3}$ & $\left(-0.382,-0.786,-0.486\right)$ \\
 \hline
 $C_{4}$ & $\left(-0.382,-0.786,0.486\right) \leftrightarrow \left(-0.382,0.786,-0.486\right)$  \\
\hline
\end{tabular}
\caption{Fixed cycles up to length two of the map $\mathcal{M}_{\mathrm{L}}$. Values are rounded to three decimal places.}
\label{tableM}
\end{table}
%%%%%%%%%%%%%%%%%%%%%%%%%%%%%%%%%%%%%%%%%%%%%%%%%%%%%%%%%%%%%%%%%%%%%%%%%%%%%%%%

One can also determine the fixed cycles of the transformation ${\mathcal M}_{\mathrm{L}}$ up to length two (see Table \ref{tableM}), and these can be proven to be equal to those of $f_{\mathrm{L}}$ apart from the single fixed point $C_{0}$, which corresponds to the maximally mixed state. The other cycles $C_{j}$ ($j=1..4$) each correspond to the repelling pure-state cycles $c_{j}$. The analytical determination of longer fixed cycles is very hard, therefore, in what follows we will rely on numerical methods to show that the map ${\mathcal M}_{\mathrm{L}}$ does not have any other mixed-state fixed cycle apart from $C_{0}$. Furthermore, by performing a Taylor-series expansion of the map up to order three around the fixed point $C_{0}$, it can be shown that $C_{0}$ is an attractive fixed point of the map from all directions.

\section{Dynamics for noisy inputs}\label{dyn}
\begin{figure}[ht]
\bc \includegraphics[width=9.6cm]{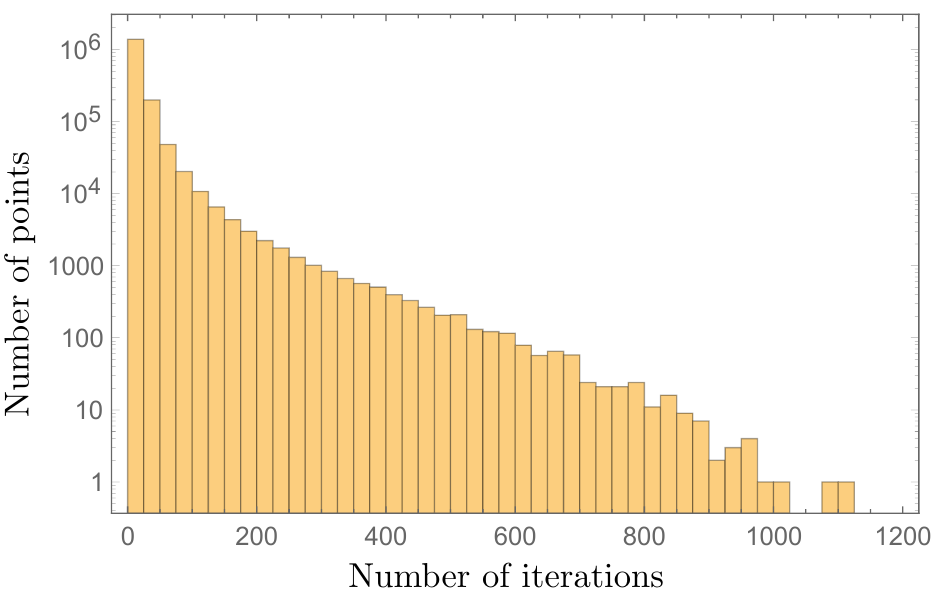}
 \ec
\vspace{-4mm}
\caption{Number of random initial states that reach $C_{0}$ with a precision of $\varepsilon=10^{-3}$ after a certain number of iterations. The total number of random initial states was $1.6\times 10^{6}$, and they were randomly chosen from a sphere of radius $1-\varepsilon$ within the Bloch sphere. %determined by the requirement that $5000$ randomly chosen points originate from the very thin spherical shell with purity $1-\varepsilon\leq P \leq 1-2\varepsilon$, with $\varepsilon=5\cdot 10^{-4}$.
}
\label{Fig1}
\end{figure}

\pst
In order to study the dynamics for noisy initial states, we performed the numerical iteration of the nonlinear map $\mathcal{M}_{\mathrm{L}}$ for randomly chosen input density matrices (uniformly distributed according to the volume of the Bloch sphere) and counted how many iterations are needed for them to converge to the (maximally mixed) fixed point $C_{0}$. We found that any such randomly chosen density matrix approaches the maximally mixed state (with a predefined precision) after a finite number of iterations. The smaller the initial noise, the larger number of iterations are needed, but eventually, all initially mixed states get close to the maximally mixed state. This can be seen in Fig.~\ref{Fig1} which shows the number of random initial states which reach $C_{0}$ with a given precision, after a certain number of iterations. The number of steps is finite for any randomly chosen initial state, which also proves the fact that there are no other attractive mixed fixed cycles of the map inside the Bloch sphere. This method, however, does not unambiguously reveal whether there are any repelling fixed cycles which could have an effect on the dynamics. 

Nonlinear pure-state dynamics of qubits are described by complex quadratic rational maps (see Sec.\ref{nonlinmap}). The repelling fixed cycles of such maps (which are all contained in the Julia set) can be found by performing the backwards iteration of the map, i.e., the iteration of its two inverses (quadratic rational functions have two inverse transformations). This is due to the property that repelling cycles become attractive under the effect of the inverse transformation \cite{Milnorbook}. In the case of $f_{\mathrm{L}}$ backwards iteration can be used to determine the Julia set itself. 

In order to find whether the nonlinear map $\mathcal{M}_{\mathrm{L}}$ has repelling fixed cycles among mixed states, we use a similar approach: we determine the inverse transformations of the map and apply backwards iteration to see whether there is convergence to some points inside the Bloch sphere. Using Eq.~(\ref{uvw_transf}) and the condition that $-1\leq\mathcal{U}, \mathcal{V}, \mathcal{W}\leq 1$, one can determine $\left(u,v,w\right)$ as a function of $\left(\mathcal{U}, \mathcal{V}, \mathcal{W}\right)$: 
\be
u_{+,-}=\pm \sqrt{\frac{1+w^{2}}{2}\left(\mathcal{U}+\sqrt{\mathcal{U}^{2}+\mathcal{W}^{2}}\right)}, \quad v_{+,-}=-\frac{\mathcal{W}\left(1+w^{2}\right)}{2u_{+,-}}, \quad w=\frac{1-\sqrt{1-\mathcal{V}^{2}}}{\mathcal{V}}.
\ee
Thus, $\mathcal{M}_{\mathrm{L}}$ has two inverses corresponding to the two transformations given by $\left(u_{+},v_{+},w\right)$ and $\left(u_{-},v_{-},w\right)$, which we denote by $m_{\mathrm{L}}^{+}$ and $m_{\mathrm{L}}^{-}$. This means that every point has two pre-images in every step which can be determined by the two inverse transformations $m_{\mathrm{L}}^{+}$ and $m_{\mathrm{L}}^{-}$. There are two special branches of backwards iteration: $\left(m_{\mathrm{L}}^{+}\right)^{\circ n}$ and $\left(m_{\mathrm{L}}^{-}\right)^{\circ n}$. It can be shown numerically that by the application of the special branch $\left(m_{\mathrm{L}}^{+}\right)^{\circ n}$ every initial state converges to the pure fixed point $C_{1}$, while by $\left(m_{\mathrm{L}}^{-}\right)^{\circ n}$ initial states are mapped to one of the pure fixed cycles $C_{2}$, $C_{3}$, and $C_{4}$, and in this case the convergence is substantially slower.

\begin{figure}[ht]
\bc \includegraphics[width=8.6cm]{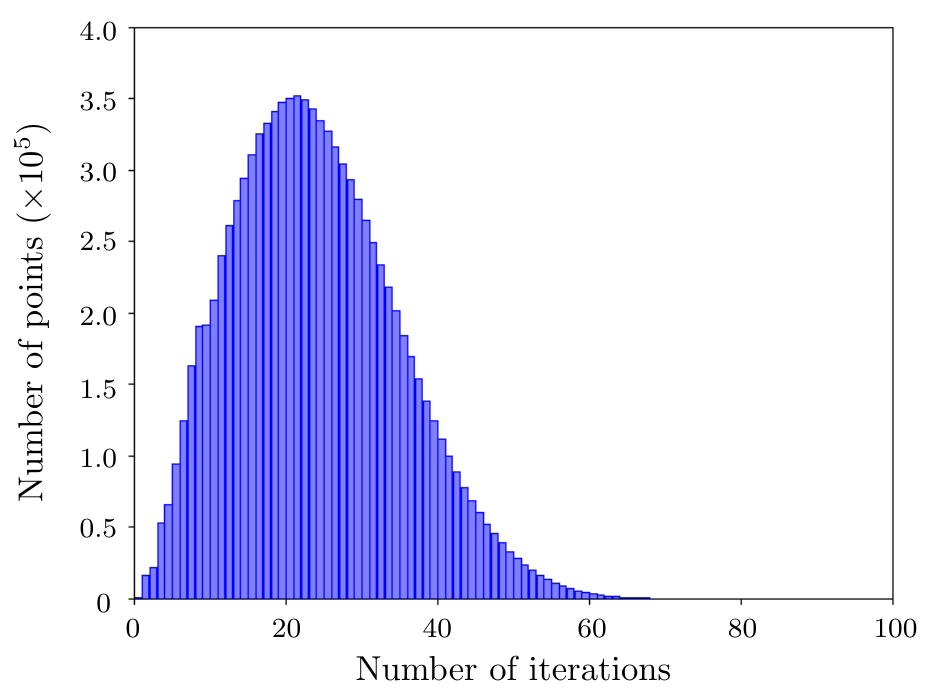}
 \ec
\vspace{-4mm}
\caption{
Number of random initial states which, when backwards iterating the map, converge to some pure state, i.e., their purity becomes $P\geq 0.99$. Uniformly distributed initial points were chosen from the spherical volume of radius $\varepsilon=10^{-2}$ around the center of the Bloch sphere ($C_{0}$). In each iteration, we randomly chose $m_{\mathrm{L}}^{+}$ or $m_{\mathrm{L}}^{-}$ to determine the pre-image of the point. }
\label{Fig2}
\end{figure}

In order to find out whether there exist fixed cycles inside the Bloch sphere which behave attractively when backwards iterating the map, we have randomly chosen points from the close neighborhood of the center of the Bloch sphere, i.e., $C_{0}$, which is repelling under the application of the inverse maps. Then, in each step, we have randomly chosen $m_{\mathrm{L}}^{+}$ or $m_{\mathrm{L}}^{-}$ to determine the pre-image of the point. We continued this process until states converged with a given precision to some pure state (i.e., to the Julia set of the pure map). Fig.~\ref{Fig2} shows the number of random initial states which reached a pure state after a certain number of iterations. We have not found any initial states that would not reach some pure state after a finite number of steps. Furthermore, increasing the radius of the spherical volume of initial states around $C_{0}$ did not indicate otherwise. Therefore we may assume that the nonlinear map $\mathcal{M}_{\mathrm{L}}$ does not have any repelling fixed point inside the Bloch sphere.  

From a practical point of view, our results show that the nonlinear protocol of the Latt\`{e}s map is very sensitive to initial noise. Even though for pure states, the transformation results in chaotic dynamics, this property is eventually destroyed by the presence of noise in the ensemble state. However, our numerical results indicate (see Fig.~\ref{Fig1}) that there may be initial states which need more than $1000$ steps to converge to the maximally mixed state, and that these originate from the close vicinity of pure states (i.e., only very small noise is present in the ensembe state). It is an open question whether the property of exponential sensitivity, which was proven for the pure-state dynamics, remains valid in the case of these close-to-pure states up to a certain number of iterations.

\section{Summary}
\pst
We have shown that the property of chaotic evolution which is present in the case of pure states under the application of the nonlinear quantum protocol described a Latt\`{e}s-type of map, is sensitive to noise in the initial ensemble. This property is so far unlike to other nonlinear protocols where the most relevant characteristics of the dynamics are clearly preserved in the presence of some initial noise \cite{Kiss2011,TBAKK2017,KK2018,Martin}. Moreover, although the so-far studied nonlinear protocols can be used for quantum state purification \cite{Kiss2011,Martin} or even quantum state discrimination \cite{TBAKK2017,KK2018} this Latt\`{e}s-type of protocol does not possess any such property. It is plausible to assume that the sensitivity to initial noise is a general property of Latt\`{e}s-type of maps, where the Julia set is identical to the set of all pure states.

\section*{Acknowledgments}
\pst
T. K. and O. K. are grateful for the support of the National Research, Development and Innovation Office of Hungary (Project Nos. K115624, K124351, PD120975, 2017-1.2.1-NKP-2017-00001) and the Lendulet Program of the Hungarian Academy of Sciences (LP2011-016). O. K. acknowledges support from the J\'{a}nos Bolyai Research Scholarship of the Hungarian Academy of Sciences.  I. J. received support from the Czech Grant Agency under grant No. GA \v CR 16-09824S and from MSMT RVO 68407700. This publication was funded by the project "Centre for Advanced Applied Sciences", Registry No. CZ.$02.1.01/0.0/0.0/16\_019/0000778$, supported by the Operational Programme Research, Development and Education, co-financed by the European Structural and Investment Funds and the state budget of the Czech Republic.

\end{document}